\documentclass{llncs}
\usepackage{mathptmx}       
\usepackage{helvet}         
\usepackage{courier}        
\usepackage{type1cm}        
\usepackage{makeidx}         
\usepackage{graphicx}        
\usepackage{multicol}        
\usepackage[bottom]{footmisc}
\usepackage{subfigure}
\usepackage{amsfonts}
\usepackage[cmex10]{amsmath}

\usepackage{algorithm}
\usepackage{algpseudocode}

\usepackage{hyperref}       
\usepackage{url}            
\usepackage{booktabs}       
\usepackage[utf8]{inputenc}
\usepackage[misc]{ifsym}
\usepackage{comment}
\usepackage{multirow}
\usepackage{dirtytalk}
\usepackage{mathtools}
\usepackage{amsmath}

\begin{document}

\title{icon: Fast Simulation of Epidemics on Coevolving Networks}
\titlerunning{Fast Simulation of Epidemics on Coevolving Networks}  
%
\author{Gerrit Großmann, Sebastian Vollmer\phantom{\inst{1}}}
\authorrunning{Großmann et al.} 
\institute{German Research Center for Artificial Intelligence (DFKI)\\ Data Science and its Applications (DSA) Research Group\\ Kaiserslautern, Germany\\
\email{gerrit.grossmann@dfki.de}
}

\maketitle

\begin{figure}[h!]
\centering
\includegraphics[width=0.7\textwidth]{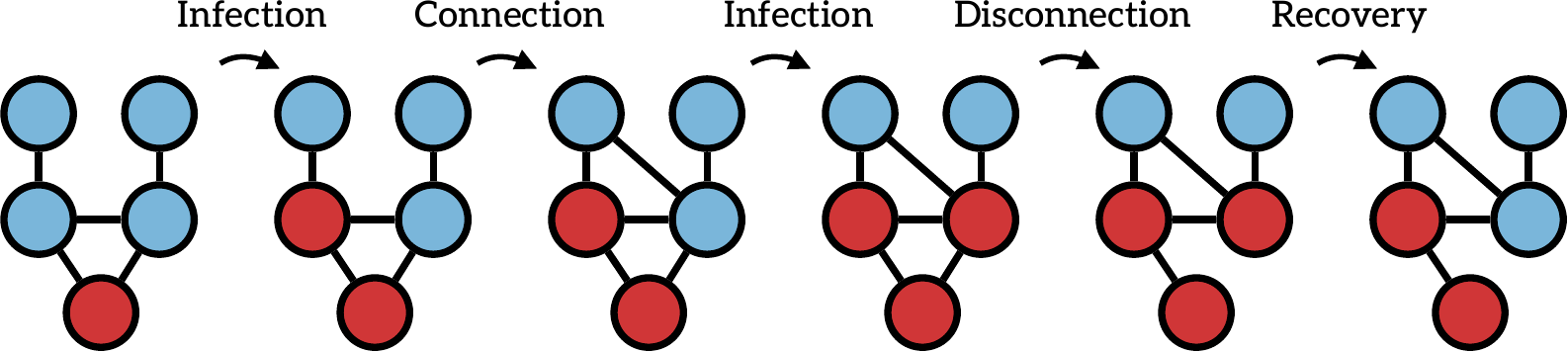}
\caption{Schematic illustration of our continuous-time coevolving spreading model inspired by \cite{wang2019adaptive}. Similar to the standard SIS model, infected (red) nodes can 1) transmit their infection to neighboring susceptible (blue) nodes and can 2) recover (become susceptible again). Additionally, two connected infected nodes can dissociate (removing their edge), and two unconnected susceptible nodes can associate (creating an edge between them).}
\label{fig:plots}
\end{figure}

\begin{abstract}
We introduce a fast simulation technique for modeling epidemics on adaptive networks. Our rejection-based algorithm efficiently simulates the co-evolution of the network structure and the epidemic dynamics. We extend the classical SIS model by incorporating stochastic rules that allow for the association of susceptible nodes and the dissociation of infected nodes. The method outperforms standard baselines in terms of computational efficiency while revealing new emergent patterns in epidemic spread.
\end{abstract}

\section{Introduction}
\noindent 
The assumption that epidemics occur on static networks is one of the most drastic abstractions in epidemic modeling. While this simplification allows for significant reduction in complexity, real-world networks are typically dynamic, and accounting for this dramatically alters the behavior of the model \cite{bansal2010dynamic,persoons2024transition}. 

This work introduces a rapid simulation technique for epidemics on adaptive networks, aiming to inspire further research into these complex systems.
We propose \textsf{icon}, a rejection-based simulation algorithm for \textsf{i}nfectious disease models on \textsf{co}-evolving \textsf{n}etworks.
In \emph{coevolving} (or \emph{adaptive}, a special case of \emph{temporal} or \emph{time-varying} \cite{masuda2016guide,masuda2017introduction}) networks, the network structure evolves in response to the spreading dynamics and vice versa. We extend the classical stochastic and continuous-time SIS (Susceptible-Infected-Susceptible)  model by incorporating rules that allow for the association (creating an edge) of two unconnected susceptible nodes and the dissociation (removing an edge) of connected infected nodes (cf.\ Figure~\ref{fig:plots}).

We demonstrate that our approach outperforms rejection-free baselines by a wide margin (cf.\ Figure~\ref{fig:results}) and that it reveals interesting emergent patterns, highlighting the need for further investigation (cf.\ Appendix~\ref{sec:results-varying-parameters}).

\paragraph{Code Availability.}
Code is made available at {\hyperlink{https://github.com/GerritGr/icon}{\small \texttt{github.com/GerritGr/icon}}}.

\paragraph{Model.}
We consider a contact graph where each node is in one of two states (\emph{susceptible} (S) or \emph{infected} (I)) at each point in time. Following exponentially distributed waiting times, infected nodes infect their susceptible neighbors at rate $\beta$ and recover at rate $\alpha$. Additionally, each pair of connected infected nodes can remove their edge at rate $b$, while two unconnected susceptible nodes can create an edge between them at rate $a$. 
This process gives rise to a continuous-time Markov chain (CTMC) semantics \cite{kiss2017mathematics}.
Our model is influenced by the one proposed in \cite{wang2019adaptive}, but instead of breaking SI edges, we break II edges, a mechanism that, to our knowledge, has not been studied before. 

\paragraph{Related Work.}
The study of epidemics on coevolving networks is a vast area, and we refer the interested reader to \cite{masuda2016guide,masuda2017introduction}. Many models of how network structure and epidemic coevolve have been proposed. Alternative mechanisms include rewiring based on population awareness \cite{gross2008robust,zhang2021coevolving}, neighborhood exchange \cite{volz2007susceptible}, assortative mixing \cite{newman2002assortative}, population growth \cite{demirel2017dynamics}, or by replacing infected neighbors with susceptible ones \cite{marceau2010adaptive}.

Most existing work on fast simulation techniques considers static networks in both the Markovian \cite{cota2017optimized,grossmann2019rejection,st2019efficient} and non-Markovian \cite{boguna2014simulating,masuda2018gillespie,grossmann2019rejection,pelissier2022practical} settings. Work on temporal networks primarily focuses on network dynamics driven by external processes \cite{vestergaard2015temporal,ahmad2019continuous,holme2021fast,hambridge2023methods}. 
The existing methods we found in the literature \cite{kiss2017mapping,rocha2013epidemics,sayama_simulating} do not use rejection-based techniques, and none of them have been tested systematically. Additionally, they are often highly specific, runtime results are typically not reported, and implementations are not optimized for runtime efficiency.

\section{Our Method: \textsf{icon}}
\begin{figure}[t!]
\centering
\includegraphics[width=0.9\textwidth]{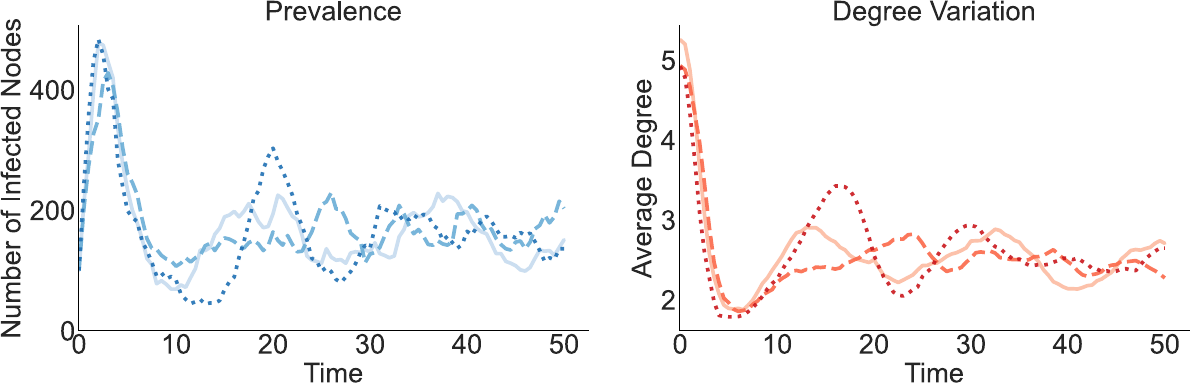}
\caption{Evolution of prevalence (left) and average degree (right) for three example trajectories on a Erdős–Rényi graph model with 1000 nodes. }
\label{fig:traces}
\end{figure}

Our rejection-based simulation method is conceptually simple. The main data structures are a dictionary (or hashmap) that maps each node to its current state ({S} or {I}) and a list storing all edges in the current graph. At each simulation step, we first determine whether the next event occurs at the node level (i.e., recovery), at the edge level (i.e., transmission or disconnection), or at the level of unconnected nodes (i.e., two susceptible nodes form a new connection).
By over-approximating the relevant rates, this can be done in constant time. Specifically, we assume that all nodes are infected to calculate the rate of node-level events. Similarly, we assume that each edge is an SI edge (or an II edge if $\beta < b$) to overestimate the rate of edge events, and that all nodes are disconnected and susceptible to overestimate the rate of new edge formations. This event computation is followed by a rejection step to correct for the over-approximation. While the number of rejection steps can be large, since each rejection is computed in constant time (i.e., independent of the number of nodes/edges), the method scales extremely well with network size.

Our method can be extended in various ways, for example, to cases where the edge creation probability is not uniform across all node pairs (e.g., it may depend on the shortest path distance), or where not only II edges break but also SI edges, or even where awareness spreads through a different network layer.
Correctness (statistical equivalence to the baselines) of our approach follows from the same principles as in \cite{grossmann2019rejection,cota2017optimized}. Pseudocode is provided in Appendix \ref{sec:icon}.

\section{Experiments}
\begin{figure}[t!]
\centering
\includegraphics[width=\textwidth]{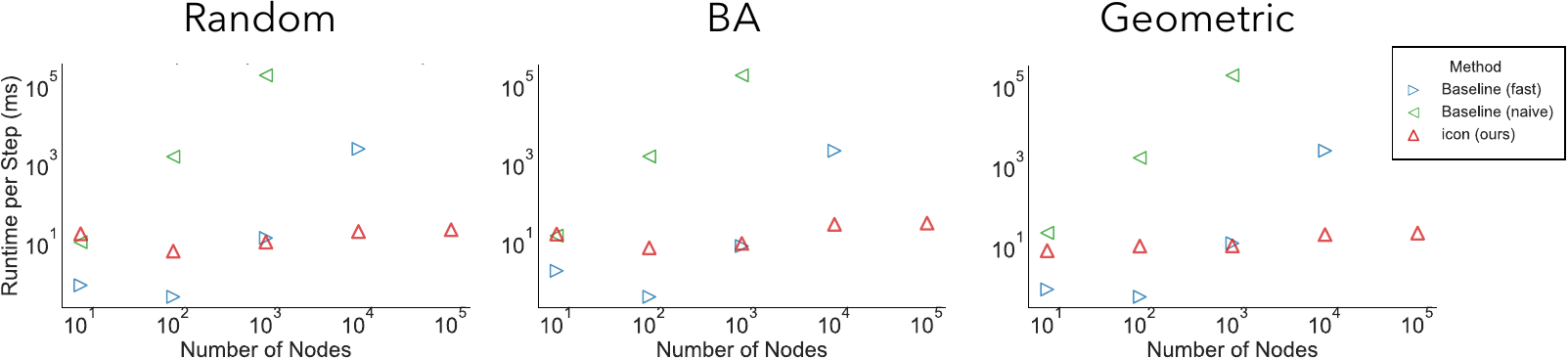}
\caption{[Lower is better.] Mean CPU time of our rejection-based method (icon) compared to two baselines using three graph models based on five runs (see also Appendix~\ref{sec:detailed-runtime-results}). }
\label{fig:results}
\end{figure}

\paragraph{Setup.}
We test one parameterization of our model on three different network types (random, Barabási–Albert, geometric) and measure how the CPU runtime for each step (counting only accepted steps) increases with the number of nodes. Example dynamics are shown in Figure~\ref{fig:traces}. Details of the experimental setup are provided in Appendix~\ref{sec:experiment-details}.

\paragraph{Baseline.}
We test two baselines. The first is a \emph{naive} method that works as follows: For each step, all pairs of nodes are considered. If a pair can potentially react, we generate a random waiting time. Additionally, we generate a random waiting time for each infected node to recover. Finally, the smallest waiting time is selected, and the corresponding event is applied. For the \emph{fast} baseline, we aimed at building the fastest possible method without rejections. It circumvents the costly iteration over all node pairs using lists of edges of specific types. More details are provided in Appendix~\ref{sec:baseline2}.

\paragraph{Results.}
The results are shown in Figure~\ref{fig:results}, with additional details provided in Appendix~\ref{sec:detailed-runtime-results}. Our method clearly scales much better with network size compared to the baselines. Trajectories for for different parameters are reported in Appendix \ref{sec:results-varying-parameters}.

\section{Conclusions and Future Work}
With this work, we aim to contribute to the study of the intricate dynamics of coevolving systems. 
In addition to studying how the degree distribution and prevalence change, it would certainly also be interesting to explore the temporal evolution of more epidemic-related indicators, such as the effective reproduction number ($R_t$) or the epidemic threshold.
For future work, we also plan to investigate scenarios where properties of the global network structure (e.g., degree distribution or community structure) remain unchanged to enforce more realistic conditions. Additionally, we are working on a Julia implementation of our method and explore event-based methods.

\bibliographystyle{splncs03} 
\bibliography{refs} 

\appendix 

\section{Detailed Description of \textsf{icon}}
\label{sec:icon}

We present \textsf{icon} in Algorithm \ref{alg:icon}.
Note that by selecting appropriate data structures for $E$ and $M$, all operations (insertion, deletion, random selection, identifying the length) can be performed in (amortized) constant time with respect to the number of nodes and edges. However, the number of rejection steps might increase with the size of the network, particularly in regions with low infection counts.
\begin{algorithm}
\caption{\textsf{icon}: Rejection-based Simulation for Coevolving Epidemics on Networks}
\label{alg:icon}
\begin{algorithmic}[1]
\State \textbf{Input:}
\State \quad Recovery rate $\alpha \in \mathbb{R}_{>0}$
\State \quad Infection rate $\beta \in \mathbb{R}_{>0}$
\State \quad Connection rate $a \in \mathbb{R}_{\geq 0}$
\State \quad Disconnection rate $b \in \mathbb{R}_{\geq 0}$
\State \quad Time horizon $H \in \mathbb{R}_{>0}$
\State \quad Initial contact graph $\mathcal{G} = (\mathcal{V}, \mathcal{E})$
\State \quad Initial node states $L: \mathcal{V} \rightarrow \{\texttt{I}, \texttt{S}\}$
\State \textbf{Output:}
\State \quad Sequence of events
\State Initialize time $t = 0$
\State Initialize $n = |\mathcal{V}|$
\State Extract edgelist $E$ from $\mathcal{G}$ such that $v_1 < v_2$ for all $(v_1, v_2) \in E$
\While{$t < H$}
    \State Compute rates:
    \State \quad $r^V = \alpha n$ 
    \State \quad $r^E = \max(b, \beta) \cdot |E|$ 
    \State \quad $r^D = a \cdot \frac{n(n - 1)}{2}$ 
    \State $t^V \sim \text{Exp}(r^V)$ \Comment{Sample exponentially distributed waiting times}
    \State $t^E \sim \text{Exp}(r^E)$ 
    \State $t^D \sim \text{Exp}(r^D)$ 
    \State Sample next event time $t' = \min(t^V, t^E, t^D)$
    \State Update time: $t = t + t'$
    \If{$t' = \text{Exp}(r^V)$}
        \State Select $v \in \mathcal{V}$ uniformly at random
        \If{$L(v) = \texttt{I}$}
            \State $L(v) = \texttt{S}$ \Comment{Recovery: infected node becomes susceptible}
        \EndIf
    \ElsIf{$t' = t^E$}
        \State Select $(v_1, v_2) \in E$ uniformly at random
        \State Sample $u \sim \text{Uniform}(0, 1)$
        \If{$L(v_1) = \texttt{I}$ and $L(v_2) = \texttt{I}$}
            \If{$u < \frac{b}{\max(b, \beta)}$}  \Comment{Rejection based on the upper bound.}
                \State Remove $(v_1, v_2)$ from $E$ \Comment{Disconnection of II edge}
            \EndIf
        \ElsIf{($L(v_1) = \texttt{I}$ and $L(v_2) = \texttt{S}$) or ($L(v_1) = \texttt{S}$ and $L(v_2) = \texttt{I}$)}
            \If{$u < \frac{\beta}{\max(b, \beta)}$}  \Comment{Rejection based on the upper bound.}
                \State Set $L(v_i) = \texttt{I}$ \Comment{Transmission ($v_i$ is the susceptible one of the two nodes.)}
            \EndIf
        \EndIf
    \ElsIf{$t' = t^D $}
        \State Select two different nodes $(v_1, v_2) \in \mathcal{V}$ uniformly at random ($v_1 < v_2$).
        \If{$L(v_1) = \texttt{S}$ and $L(v_2) = \texttt{S}$ and $(v_1, v_2) \notin E$}
        \State Add edge $(v_1, v_2)$ to $E$ \Comment{Connection of two susceptible nodes}
        \EndIf
    \EndIf
\EndWhile
\end{algorithmic}
\end{algorithm}

\section{Detailed Description of the Baseline}
\label{sec:baseline2}

In this section, we describe how the \say{fast} baseline works. The basic data structures are a graph $\mathcal{G} = (\mathcal{V}, \mathcal{E})$ and a mapping $L: \mathcal{V} \rightarrow \{\texttt{I}, \texttt{S}\}$ that stores the current state of each node. Additionally, we maintain four lists: one for infected nodes, one for SI edges, one for II edges, and one for pairs of susceptible nodes that are not adjacent.

At each simulation step, we use the lengths of these lists to determine the rates (and corresponding waiting times) for the four possible event types. Similar to Algorithm~\ref{alg:icon}, we then identify the event type with the shortest waiting time and apply the event to the graph. This involves sampling a random element (node or edge) from the selected list and updating the mapping $M$ accordingly.

The challenging part is updating the remaining lists after an event. For instance, when a node changes from susceptible to infected, we need to remove all occurrences of this node from the list of susceptible (unconnected) node pairs. Similarly, we must add new edges to the list of infected neighboring nodes. To avoid iterating over the entire list, we only update the pairs that are directly affected by the change in the node’s state by identifying the neighbors of the updated node and modifying only the relevant pairs in the lists.

However, the runtime of this method depends on the number of edges connected to a node, meaning it does not scale as efficiently with network size as our \textsf{icon} method.

\section{Experiment Details}
\label{sec:experiment-details}

We use the following model parameters: $\alpha = 1.0$, $\beta = \frac{3}{\langle d \rangle}$ (where $\langle d \rangle$ is the mean degree of the initial graph), $a = \frac{2.0}{n}$ (where $n$ is the number of nodes), and $b = 2.0$. By making the parameters dependent on the graph size and connectivity, we can apply the same parameter set across different networks without the risk of immediate die-out.

We initialize the simulations with $10\%$ of the nodes randomly chosen as infected. The graphs were generated using the NetworkX library. The random (Erdős–Rényi) graph has a connection probability of $p = 5 / (n - 1)$, resulting in an average degree of approximately 5. The Barabási–Albert graph was created with $m = 5$, where $m$ represents the number of edges attached from a new node to existing nodes. The geometric graphs were constructed to have an average degree of approximately 5.

The method was implemented using Python and all experiments executed on a standard desktop computer with 32 GB of RAM and an Intel i9-10850K CPU.

\section{Results on Varying Parameters}
\label{sec:results-varying-parameters}

Note that we essentially have three free parameters (four parameters; but when their relative proportions is unchanged, only the speed of the dynamics changes). Thus, following standard convention, we always keep $\alpha = 1.0$ and only change $\beta$, $b$, and $a$.

\begin{figure}[t!]
\centering
\includegraphics[width=0.98\textwidth]{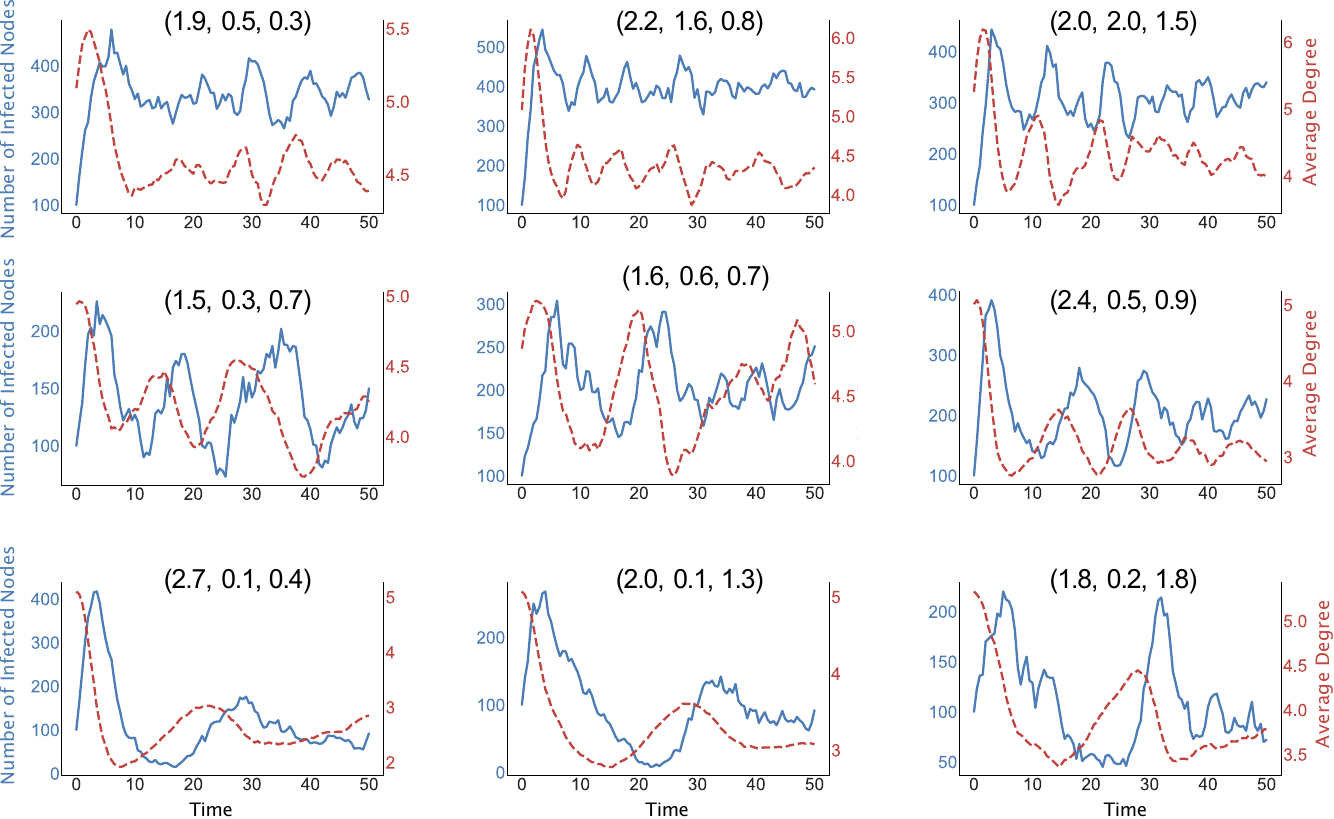}
\caption{Some results for different parameter combinations for random graphs with 1000 nodes.}
\label{fig:params}
\end{figure}

We present results for a set of random parameter combinations in Figure \ref{fig:params}. These combinations are expressed as $(\beta', a', b)$. The parameters are scaled as $\beta = \frac{\beta'}{\langle d \rangle}$, where $\langle d \rangle$ is the average degree (approximately 5 in this case), and $a = \frac{a'}{n}$, where $n$ is the total number of nodes.

As expected from mean-field approximations, we observe wave-like dynamic behaviors. Specifically, we consistently see a wave pattern in the infection dynamics, which appears largely independent of the precise parameter values. Our results show this phenomenon in a much more robust way than previously suggested by theoretical studies.

Moreover, we identify three distinct classes of behavior:
\begin{itemize}
    \item \textbf{Top row}: The mean degree almost perfectly mirrors the infection curve.
    \item \textbf{Middle row}: A growing shift between the mean degree and infection curves is observed (the minima and maxima no longer align).
    \item \textbf{Bottom row}: The system exhibits two large waves, one prominent initial wave followed by a second (typically smaller) wave. this behavior closely resembles real-world pandemics, such as COVID-19.
\end{itemize}

\section{Detailed Runtime Results}
\label{sec:detailed-runtime-results}
We document the same results as in Figure \ref{fig:results} in Table 
\ref{tab:runtime}.

\begin{table}[h!]
\centering
\caption{[Lower is better.] Mean CPU time per simulation step in ms with standard deviation. The smallest value for each combination of node number and graph type is in bold.}
\label{tab:runtime}
\begin{tabular}{@{}llcccc@{}}
\toprule
\multirow{2}{*}{Graph Type} & \multirow{2}{*}{} & \multicolumn{4}{c}{\textbf{Number of nodes}} \\ 
\cmidrule(lr){3-6}
& & $10^2$ & $10^3$ & $10^4$ & $10^5$ \\ \midrule
\multirow{3}{*}{Random} 
& icon (ours)           & 6.19 ± 0.43  & \textbf{10.52 ± 0.78 }  & \textbf{19.57 ± 0.47}  & \textbf{21.67 ± 0.19} \\
& Baseline Fast  & \textbf{0.43 ± 0.03} & {13.35 ± 0.71} & {2447 ± 73.34}  & - \\
& Baseline Naive & 1553 ± 57.37 & 181257 ± 2025.2 & - & - \\ \midrule

\multirow{3}{*}{BA} 
&  icon (ours)              & 7.22 ± 0.24  & {9.32 ± 1.04}   & \textbf{28.66 ± 2.88}  & \textbf{30.70 ± 0.39} \\
& Baseline Fast  & \textbf{0.407 ± 0.02} & \textbf{7.99 ± 0.25} & 2134.39 ± 59.09  & - \\
& Baseline Naive & 1534 ± 136.09 & 178462 ± 444.34 & - & - \\ \midrule

\multirow{3}{*}{Geometric} 
&  icon (ours)              & {10.25 ± 1.64}  & \textbf{10.40 ± 0.95}   & \textbf{19.78 ± 1.37}  & \textbf{21.73 ± 0.32} \\
& Baseline Fast  & \textbf{0.56 ± 0.16} & 12.08 ± 0.26 & 2398 ± 51.46  & - \\
& Baseline Naive & 1597 ± 86.61 & 179405 ± 913.53 & - & - \\ \bottomrule
\end{tabular}
\end{table}

\end{document}